\documentclass[preprint]{aastex631}

\received{December 2, 2022}
\revised{January 10, 2023}
\accepted{January 17, 2023}
\submitjournal{The Astrophysical Journal}

\shorttitle{Jet-Torus Interaction in NGC 1052}
\shortauthors{Kameno et al.}
\graphicspath{{./}{figures/}}

\begin{document}

\title{Probing Jet-Torus Interaction in the Radio Galaxy NGC 1052 by Sulfur-Bearing Molecules}

\correspondingauthor{Seiji Kameno}
\email{seiji.kameno@alma.cl}

\author[0000-0002-5158-0063]{Seiji Kameno}
\affiliation{Joint ALMA Observatory, Alonso de C\'{o}rdova 3107 Vitacura, Santiago 763-0355, Chile}
\affiliation{National Astronomical Observatory of Japan, 2-21-1 Osawa, Mitaka, Tokyo 181-8588, Japan}

\author[0000-0001-7719-274X]{Satoko Sawada-Satoh}
\affiliation{Graduate School of Science, Osaka Metropolitan University, 1-1 Gakuen-cho, Naka-ku, Sakai, Osaka, 599-8531, Japan}

\author[0000-0002-3443-2472]{C. M. Violette Impellizzeri}
\affiliation{Leiden Observatory, Leiden University, PO Box 9513, 2300 RA, Leiden, The Netherlands}

\author[0000-0002-4052-2394]{Kotaro Kohno}
\affiliation{Institute of Astronomy, Graduate School of Science, The University of Tokyo, 2-21-1 Osawa, Mitaka, Tokyo 181-0015, Japan}
\affiliation{Research Center for the Early Universe, Graduate School of Science, The University of Tokyo, 7-3-1 Hongo, Bunkyo-ku, Tokyo 113-0033, Japan}

\author[0000-0001-9281-2919]{Sergio Mart\'{i}n}
\affiliation{European Southern Observatory, Alonso de C\'{o}rdova 3107 Vitacura, Santiago 763-0355, Chile}
\affiliation{Joint ALMA Observatory, Alonso de C\'{o}rdova 3107 Vitacura, Santiago 763-0355, Chile}

\author[0000-0002-8726-7685]{Daniel Espada}
\affiliation{Departamento de F\'{i}sica Te\'{o}rica y del Cosmos, Campus de Fuentenueva, Universidad de Granada, E-18071 Granada, Spain}
\affiliation{Instituto Carlos I de F\'{i}sica Te\'{o}rica y Computacional, Facultad de Ciencias, E-18071, Granada, Spain}

\author[0000-0002-5461-6359]{Naomasa Nakai}
\affiliation{School of Science, Kwansei Gakuin University, 1 Gakuen Uegahara, Sanda, Hyogo 669-1330, Japan}

\author[0000-0001-6501-3871]{Hajime Sugai}
\affiliation{Environment and Energy Department, Japan Weather Association, Sunshine 60 Bldg. 55F, 3-1-1 Higashi-Ikebukuro, Toshima-ku, Tokyo 170-6055, Japan}

\author[0000-0003-1780-5481]{Yuichi Terashima}
\affiliation{Graduate School of Science and Engineering, Ehime University, 2-5 Bunkyo-cho, Matsuyama, Ehime 790-8577, Japan}

\author[0000-0002-2419-3068]{Minju M. Lee}
\affiliation{Cosmic Dawn Center (DAWN), Jagtvej 128, DK-2200 Copenhagen N, Denmark}
\affiliation{DTU-Space, Technical University of Denmark, Elektrovej 327, DK2800 Kgs. Lyngby, Denmark}

\author[0000-0003-2535-5513]{Nozomu Kawakatu}
\affiliation{Faculty of Natural Sciences, National Institute of Technology, Kure College, 2-2-11 Agaminami, Kure, Hiroshima 737-8506, Japan}



\begin{abstract}
The radio galaxy NGC 1052 casts absorption features of sulfur-bearing molecules, H$_2$S, SO, SO$_2$, and CS toward the radio continuum emission from the core and jets.
Using ALMA, we have measured the equivalent widths of SO absorption features in multiple transitions and determined the temperatures of $344 \pm 43$ K and $26 \pm 4$ K in sub-millimeter and millimeter wavelengths, respectively.
Since sub-mm and mm continuum represents the core and jets, the high and low temperatures of the absorbers imply warm environment in the molecular torus and cooler downstream flows.
The high temperature in the torus is consistent with the presence of 22-GHz H$_2$O maser emission, vibrationally excited HCN and HCO$^+$ absorption lines, and sulfur-bearing molecules in gas phase released from dust.
The origin of the sulfur-bearing gas is ascribed to evaporation of icy dust component through jet-torus interaction.
Shock heating is the sole plausible mechanism to maintain such high temperature of gas and dust in the torus.
Implication of jet-torus interaction also supports collimation of the sub-relativistic jets by gas pressure of the torus.

\end{abstract}


\section{Introduction} \label{sec:intro}
A dusty molecular torus in active galactic nuclei (AGNs) resides in the central parsec scale \citep{2016ApJ...823L..12G,2016ApJ...829L...7G,2019ApJ...884L..28I} and is recognized as a key component that plays significant roles of mass accretion onto the central engine, collimation of jets, and diversity of appearance depending on viewing angles \citep{1993ARA&A..31..473A}.
Observational studies of AGN tori are essential to understand the nature of AGNs, mass accretion processes, and jet collimation mechanisms.

The radio galaxy NGC 1052 is a unique target to probe an AGN torus which is seen nearly edge-on hiding the nucleus \citep{2020ApJ...895...73K}.
The torus harbours H$_2$O masers \citep{1994ApJ...437L..99B,1998ApJ...500L.129C,2003ApJS..146..249B,2005ApJ...620..145K} that allows sub-pc-scale kinetic studies of molecular gas with a VLBI resolution \citep{2008ApJ...680..191S}.
Presence of 22-GHz H$_2$O maser requires excitation under hot ($\sim 400$ K) and dense ($\sim 10^7$ cm$^{-3}$) condition of the molecular gas \citep{1989ApJ...346..983E}.
The torus casts various absorption features of plasma free--free absorption \citep{2001PASJ...53..169K, 2003PASA...20..134K, 2004A&A...426..481K}, dust obscuration and scattering \citep{1999ApJ...515L..61B}, 
photoelectric absorption \citep{2021ApJ...916...90B}, 
H~$\textsc{i}$ \citep{2003A&A...401..113V}, 
and molecules of OH, HCO$^{+}$, HCN, CO, SO, SO$_2$, CS, CN, and H$_2$O \citep{2002A&A...381L..29O,2004A&A...428..445L,2008evn..confE..33I,2016ApJ...830L...3S,2019ApJ...872L..21S,2020ApJ...895...73K}.
Presence of vibrationally excited HCN and HCO$^+$ absorption \citep{2020ApJ...895...73K} implies pumping by infrared radiation from warm dust \citep{2007ApJ...659..296L}. 

NGC 1052 is known to have two-sided jets with a sub-relativistic bulk speed of $0.26c - 0.53c$ \citep{2003A&A...401..113V}.
Multifrequency VLBI observations have revealed core and double-sided jet structure of NGC 1052.
The core shows a peaked flat spectrum at higher frequencies than 43 GHz \citep{2016A&A...593A..47B, 2016ApJ...830L...3S, 2019ApJ...872L..21S}, while it is optically thick at lower frequencies due to free--free absorption to hide in a gap \citep{1999AAS...194.2002K, 2001PASJ...53..169K, 2003PASA...20..134K, 2003A&A...401..113V, 2004A&A...426..481K}.
The jets show a steep spectrum implying optically thin synchrotron emission.
The brightness of the jets decreases as they flow downstreams \citep{2001PASJ...53..169K, 2003PASA...20..134K, 2003A&A...401..113V, 2004A&A...426..481K, 2016A&A...593A..47B, 2019A&A...623A..27B}.
The jet is collimated in a cylindrical shape with the width of $1.3\times 10^3\ R_{\rm S}$  (Schwarzschild radii) inside the break point of $10^4 \ R_{\rm S}$ from the core, and transit to a conical structure outside the break point \citep{2020AJ....159...14N}. \cite{2022A&A...658A.119B} found more complex upstream structure, neither cylindrical nor parabolic shape, with the second break point at $3\times 10^3 \ R_{\rm S}$.
These studies on the jet width imply collimation by surrounding dense material such as the molecular torus.

If the torus is confining the jet width by external gas pressure, we expect interaction between sub-relativistic jets and torus gas, and shocks generated by the interaction.
\cite{2019A&A...629A...4F} modelled over-pressured jets in a decreasing pressure ambient medium for NGC 1052 and simulated VLBI images reproducing the core-jet structure at 43 GHz and the gap at 22 GHz. The model indicates that pressure mismatch between the jet and the ambient medium forms recollimation shock at the nozzle to pinch the jet boundary.
Observational follow-up is desired to probe such jet-torus interaction.

Sulfur-bearing molecules are known to be good shock tracers in interstellar clouds \citep{1993MNRAS.262..915P}, young stellar objects \citep{1997ApJ...487L..93B}, protoplanetary disks \citep{2020IAUS..345..360N}, evolved stars \citep{2013ApJ...778...22A}, starburst galaxies \citep{2003A&A...411L.465M, 2005ApJ...620..210M,2007ApJ...661L.135M} and ultra-luminous infrared galaxies \citep{2022A&A...660A..82S}.
Observations of H$_2$S, SO, SO$_2$, and CS in the molecular torus of NGC 1052 would bring us a clue about jet-torus interaction.

In this article we report investigations of sulfur-bearing molecular absorption lines toward the radio continuum of NGC 1052. Our primary aim is to measure the temperature of the gas by comparing line equivalent widths (EWs) at various excitation levels. Then we will examine the jet-torus interaction as a heat source of the gas and dust in the torus.

We employ the systemic velocity of $V_{\rm LSR, radio} = 1492$ km s$^{-1}$ or $V_{\rm hel} = 1505$ km s$^{-1}$ \citep{2020ApJ...895...73K}, the bolometric luminosity of $L_{\rm bol} = 10^{42.3}$ erg s$^{-1} = 2\times 10^{35}$ W \citep{2014JApA...35..223G}, and the black hole mass of $M_{\rm BH} = 1.5 \times 10^8$ M$_{\odot}$ \citep{2002ApJ...579..530W}. The Schwarzschild radius will be $R_s = 3$ AU, corresponding to 0.17 $\mu$arcsec.
Data reduction scripts are available in GitHub repository\footnote{\href{https://github.com/kamenoseiji/ALMA-2016.1.00375.S}{https://github.com/kamenoseiji/ALMA-2016.1.00375.S}} together with spectral data.

\section{Observations and Results} \label{sec:observations}
We have carried out ALMA Band-3 (85.0 -- 88.8 GHz and 97.0 -- 100.4 GHz) and Band-4 (126.4 -- 130.1 GHz and 138.3 -- 141.7 GHz) observations of NGC 1052 targeting SO, CS, SiO, and HCN molecules.
Band-3 frequency is set to cover transitions SO $J_N = 2_2 - 1_1$ (permitted), HCN $J=1-0$, and SiO $J=2-1$, $v=0$ and $v=1$ in upper sideband (USB), and CS $(J=2-1)$, SO $J_N =3_2 - 3_1$ (permitted), and $4_5 - 4_4$ (forbidden) in lower sideband (LSB).
Band-4 USB covers SO $J_N =3_3 - 2_2$ (permitted), SiO $J=3-2$, $v=0, 1, 2$ while LSB does forbidden SO transitions.
Some SO$_2$ transitions reside in the frequency coverage.

J0238+1636 and J0243-0550 were used as the bandpass and phase calibrators, respectively, in both bands. We also used NGC 1052 for self calibration.
Table \ref{tab:obsLog} lists the observation log.

\begin{figure}
\gridline{\fig{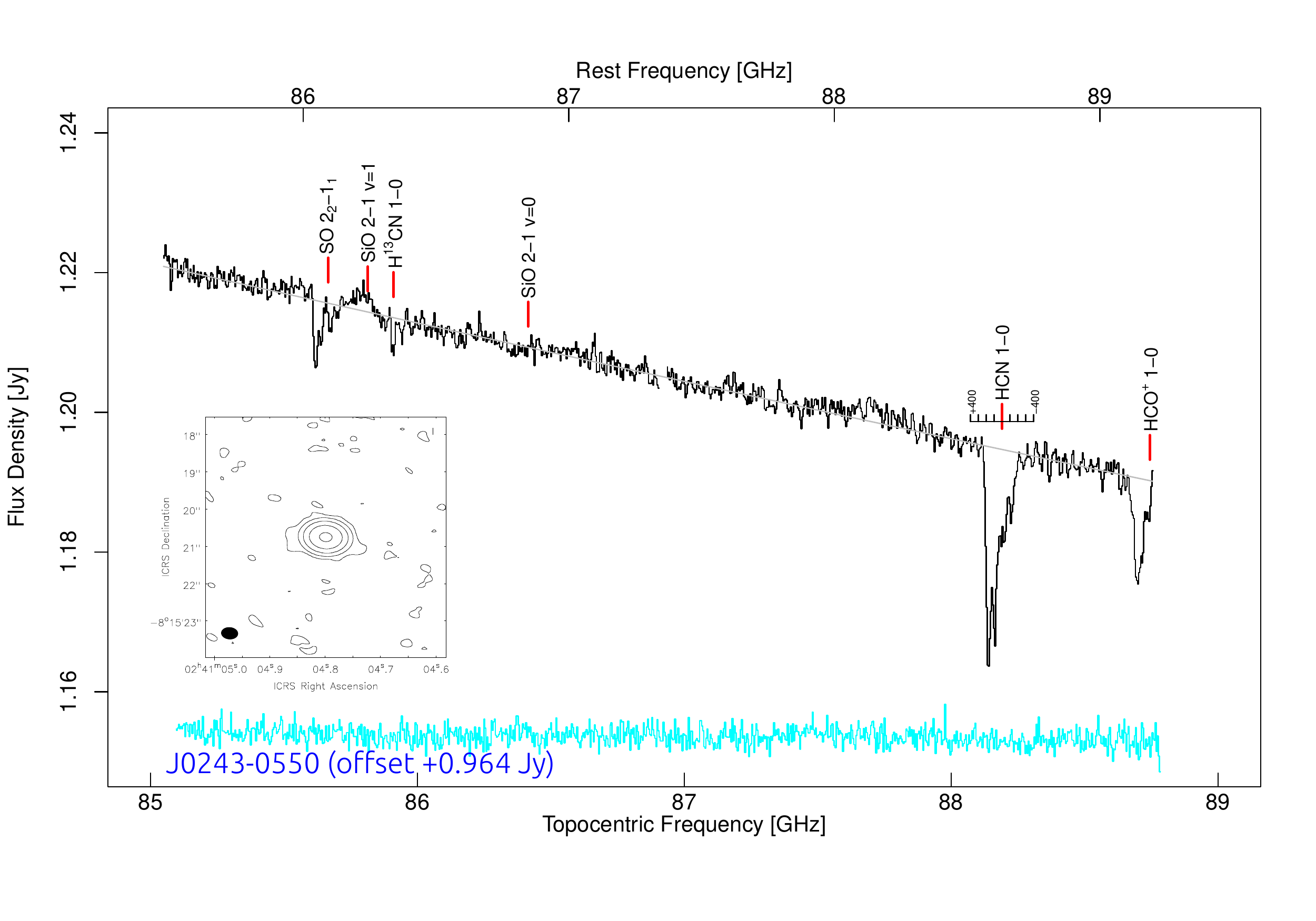}{0.475\textwidth}{(a)}
          \fig{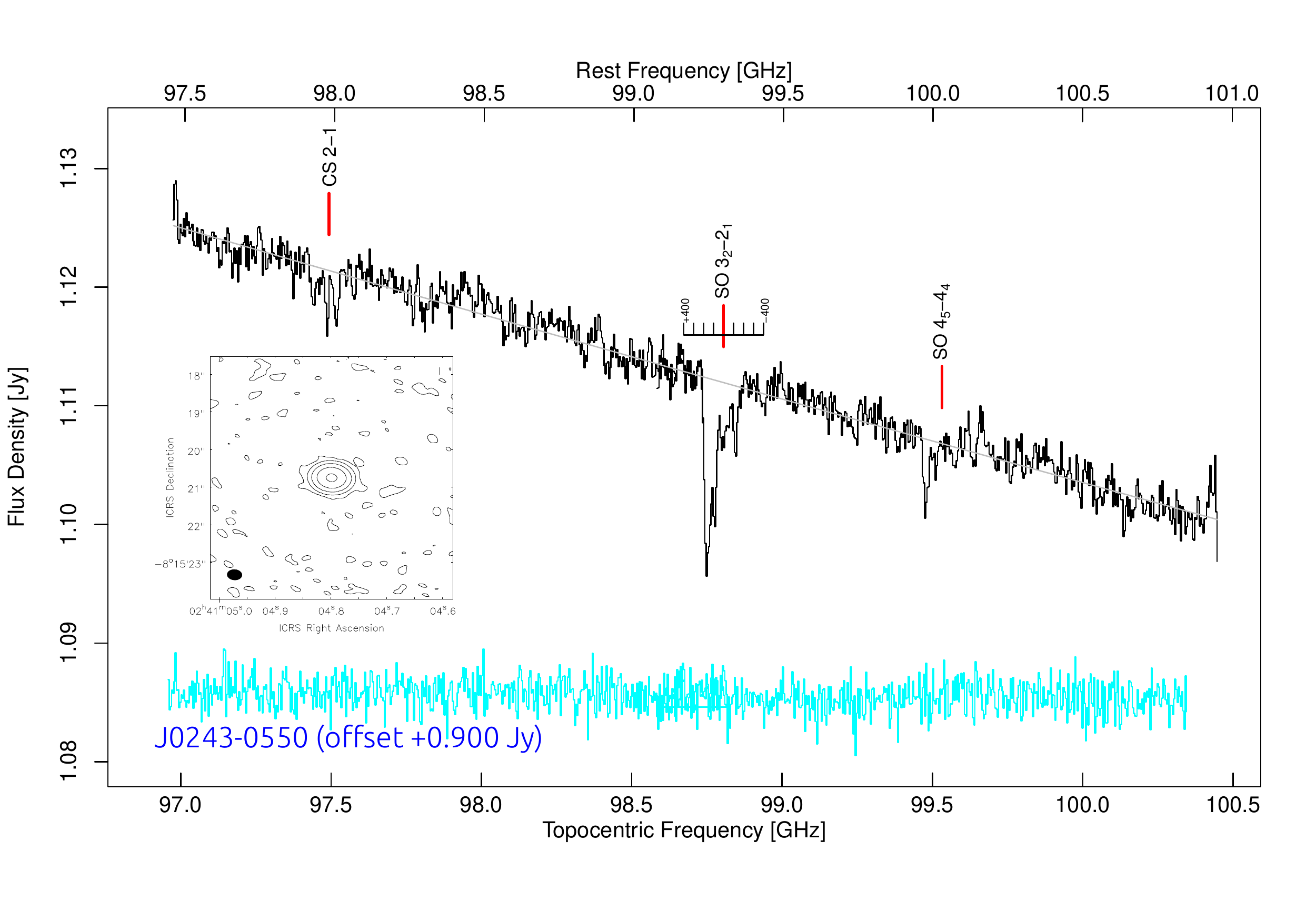}{0.475\textwidth}{(b)}}
\caption{Spectra in the single pixel at the center of NGC 1052 in Band-3. (a) Lower sideband (85.0 -- 88.8 GHz). (b) Upper sideband (97.0 -- 100.4 GHz). Top and bottom abscissas stand for the frequency at the source frame and topocentric frame, respectively. Red vertical markers indicate line species at the systemic velocity. Spectra of J0243-0550, offset by 0.964 Jy and 0.900 Jy, are shown in cyan. Continuum maps are shown inset with the contour levels of 75 $\mu$Jy beam$^{-1}$ $\times$ powers of 10.}
\label{fig:absorptionSpectrumB3}
\end{figure}

\begin{figure}
\gridline{\fig{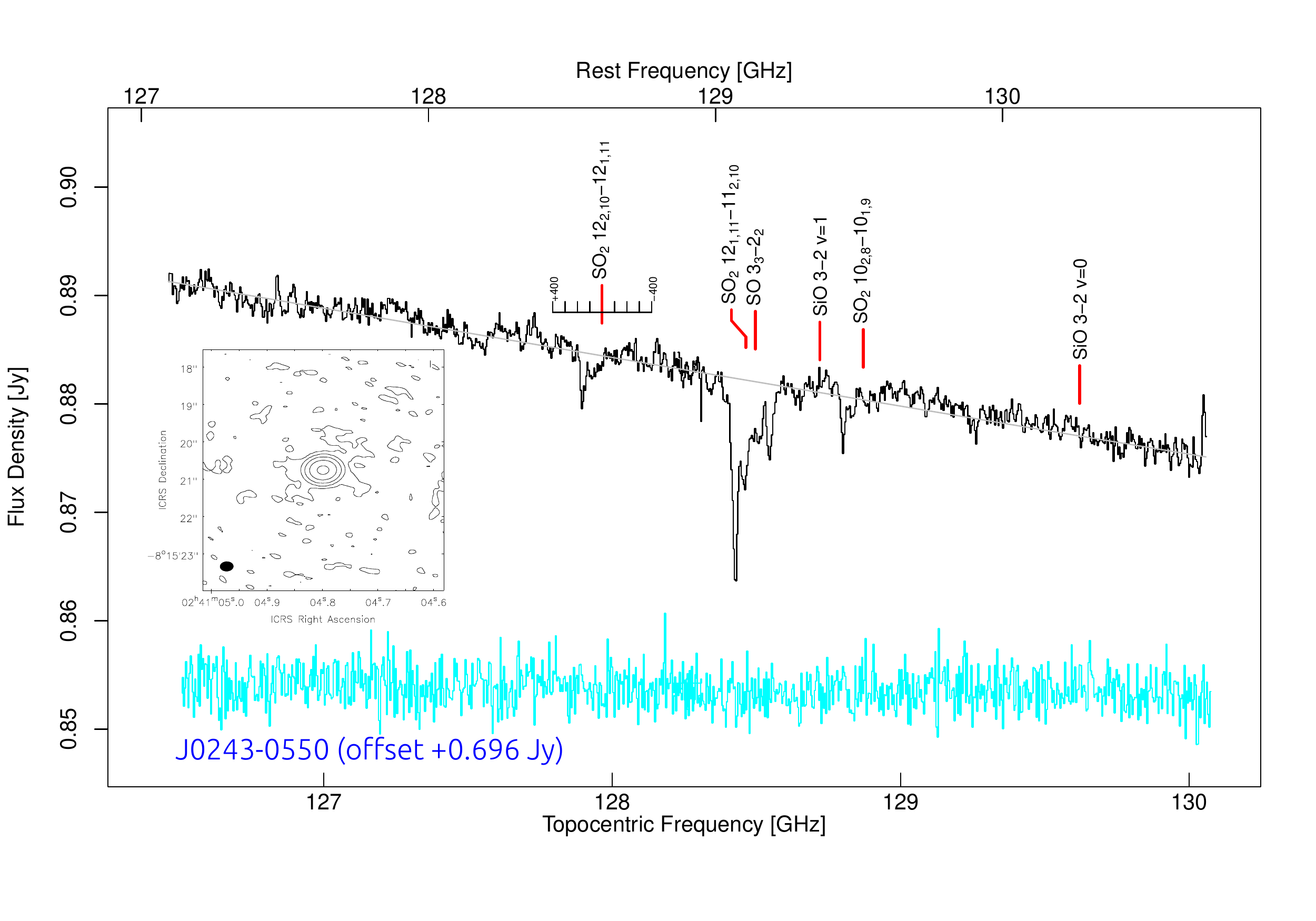}{0.475\textwidth}{(a)}
          \fig{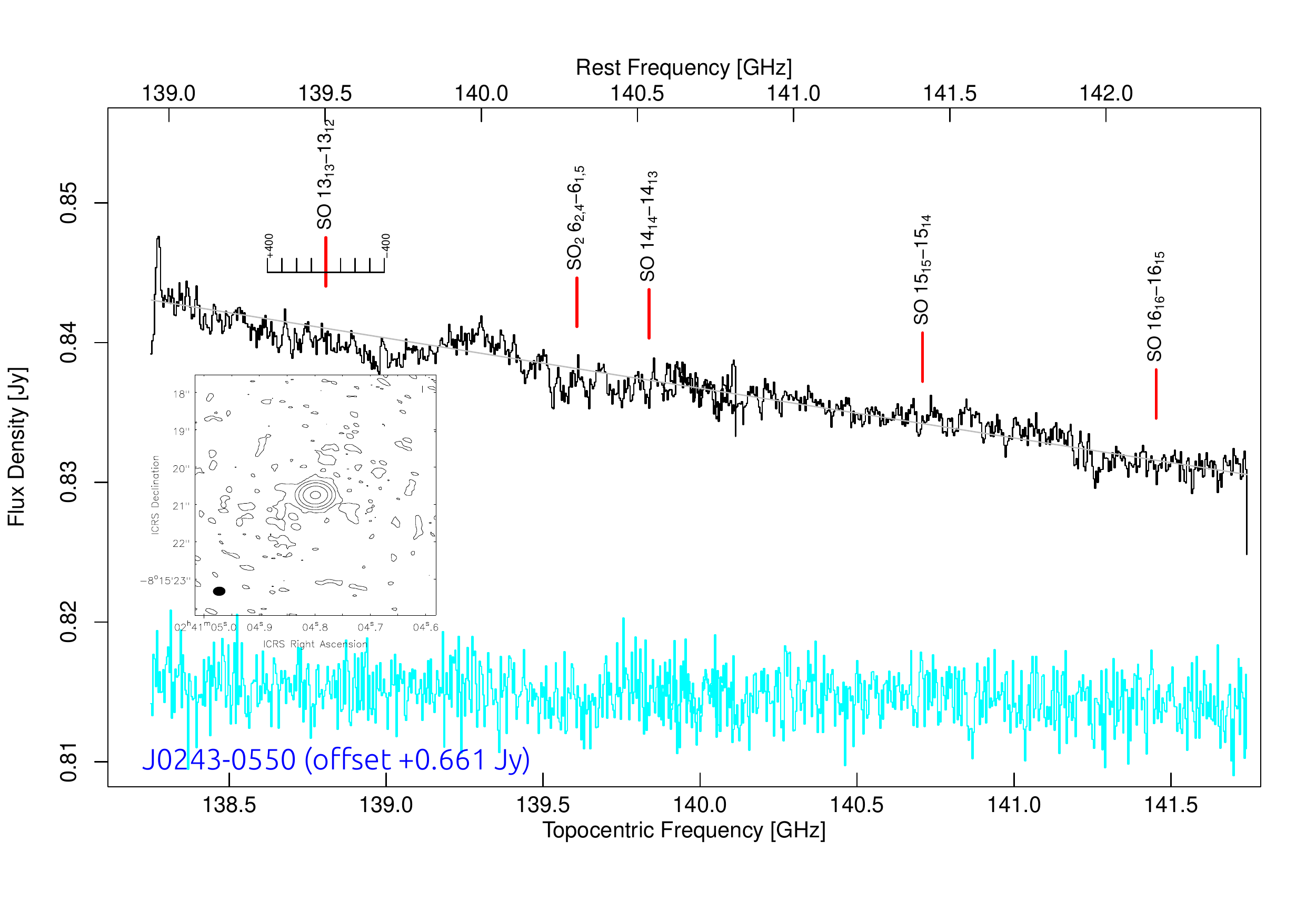}{0.475\textwidth}{(b)}}
\caption{Spectra of NGC 1052 in Band-4. (a) Lower sideband (126.4 -- 130.1 GHz). (b) Upper sideband (138.3 -- 141.7 GHz). Spectra of J0243-0550, offset by 0.696 Jy and 0.661 Jy, are shown in cyan. Continuum maps are shown inset with the contour levels of 50 $\mu$Jy beam$^{-1}$ $\times$ powers of 10.}
\label{fig:absorptionSpectrumB4}
\end{figure}

The continuum images at both bands consist of an unresolved point-like component with 1.127 Jy and 0.859 Jy at 92.7 GHz and 134.1 GHz, respectively, as shown in insets of Figures \ref{fig:absorptionSpectrumB3} and \ref{fig:absorptionSpectrumB4}.
No significant extended structure is identified with the image rms of 37 $\mu$Jy beam$^{-1}$ and 25 $\mu$Jy beam$^{-1}$, respectively.

We took spectra in a single pixel at the center of the point-like component as shown in Figures \ref{fig:absorptionSpectrumB3} and \ref{fig:absorptionSpectrumB4}.
Four plots represent LSB and USB in Band 3 and 4, respectively.
Spectra of the phase calibrator, J0243-0550, are presented in cyan as the control.
Line species are marked by red vertical ticks. Velocity scales with respect to the systemic velocity $\pm 400$ km s$^{-1}$ are also tagged on a line feature in each plot.
The identified features are listed in Table \ref{tab:absorptionLineID}.

\begin{deluxetable}{llrrrrr}
\tablecaption{Observation Log}
\tablehead{
\colhead{Date} & \colhead{ExecBlock UID} & \colhead{$\nu_{\rm obs}$} & \colhead{N$_{\rm ant}$} & \colhead{Beam size and PA} & \colhead{Spectral resolution} & \colhead{Image rms} \\
\colhead{}     & \colhead{uid://A002/}   & \colhead{GHz}              & \colhead{}             & \colhead{} & \colhead{MHz, km s$^{-1}$} &  \colhead{mJy beam$^{-1}$}
}
 \colnumbers
 \startdata
2017-07-28     & Xc2bb44/X1baf &  92.7  & 45 & $0^{\prime \prime}.42 \times 0^{\prime \prime}.30$, $+86^{\circ}$ & 3.9, 12 & 0.26 \\
2017-07-23     & Xc27dd7/X2e85 &  134.1 & 46 & $0^{\prime \prime}.34 \times 0^{\prime \prime}.25$, $-87^{\circ}$ & 3.9, 9  & 0.20 \\
\enddata
\tablecomments{(1) Year-Month-Day; (3) Center frequency; (4) Number of antennas; (5) Synthesized beam size (FWHM in major and minor axes) and position angle; (6) Spectral resolution in MHz and velocity resolution in km s$^{-1}$; (7) Image rms of channel maps
}
\end{deluxetable} \label{tab:obsLog}

HCN $J=1-0$ is the most prominent feature which shows an asymmetric profile with a sharp edge in the redshifted side.
H$^{13}$CN $J=1-0$ absorption is also detected with a more peaky profile than that of the H$^{12}$CN feature.
HCO$^+$ $J=1-0$ feature is partially covered with our spectral setting. The ratio of HCN-to-HCO$^+$ peak opacity is $2.1 \pm 0.2$, being consistent with the Korean VLBI Network (KVN) results \citep{2019ApJ...872L..21S}.

All permitted SO absorption features, $J_N = 2_2 - 1_1, \ 3_2 - 2_1,$ and $3_3 - 2_2$ are clearly detected.
SO $J_N = 4_5 - 4_4$ is the sole forbidden line which is clearly detected.
The $8_8 - 7_8$ feature is marginal and higher-$J$ forbidden lines are not detected.

We have also identified SO$_2$ absorption features of $J_{Ka, Kc} = 10_{2, 8} - 10_{1, 9}$ and $12_{2, 10} - 12_{1, 11}$.
The feature of $12_{2, 10} - 12_{1, 11}$ transition overlaps with SO $J_N = 3_3 = 2_2$. The $6_{2,4} - 6_{1,5}$ feature is not significantly detected.
CS $J=2-1$ absorption feature is marginally detected with the significance of 3.6 times the rms level.
No significant SiO absorption was detected. Possible SiO $J=2-1$, $v=1$ emission with $4.4 \pm 0.26$ mJy appeared at 56.7 km s$^{-1}$. We don't argue this feature before confirmation in another transition.

\begin{figure}
\gridline{\fig{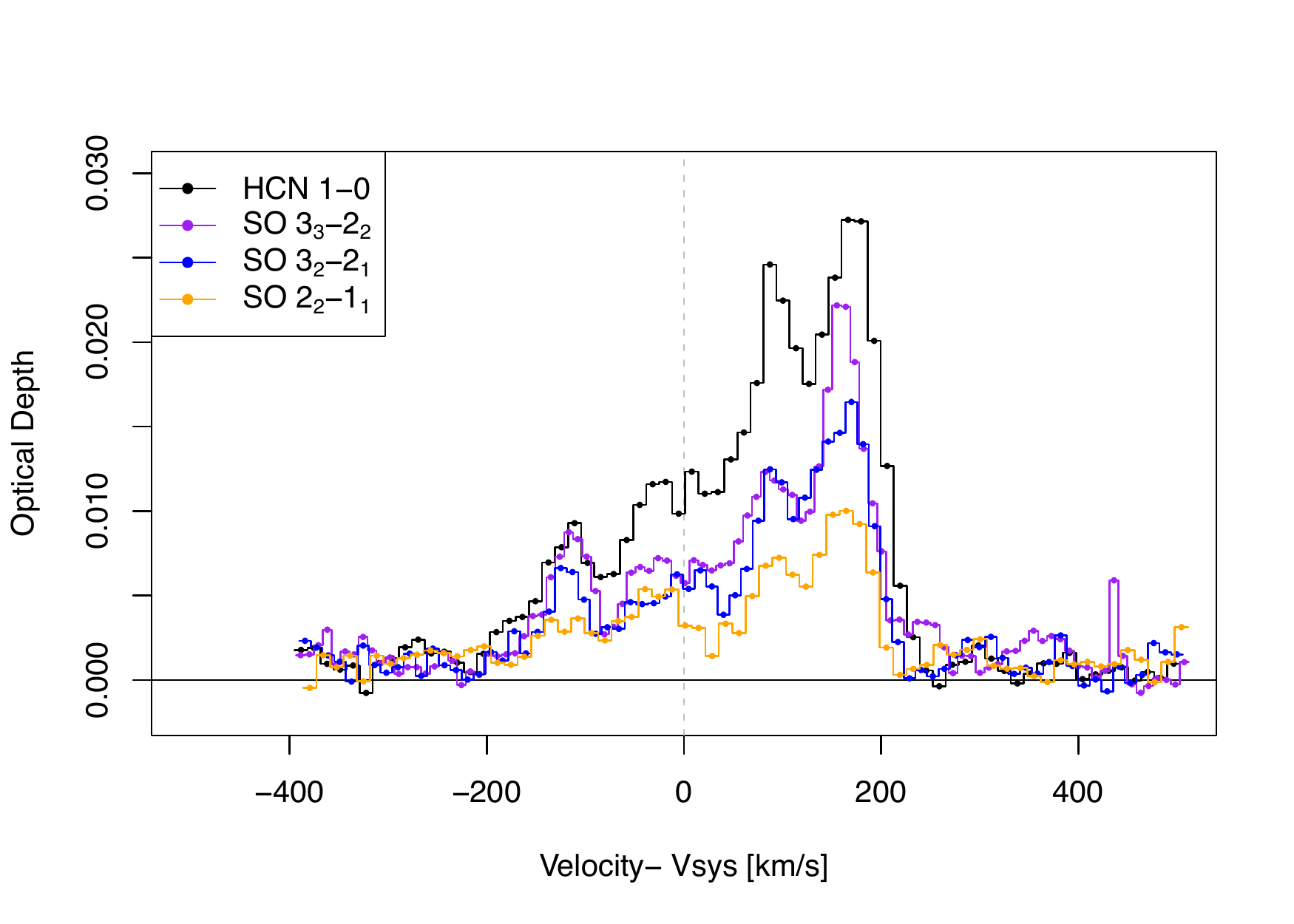}{0.475\textwidth}{(a)}
          \fig{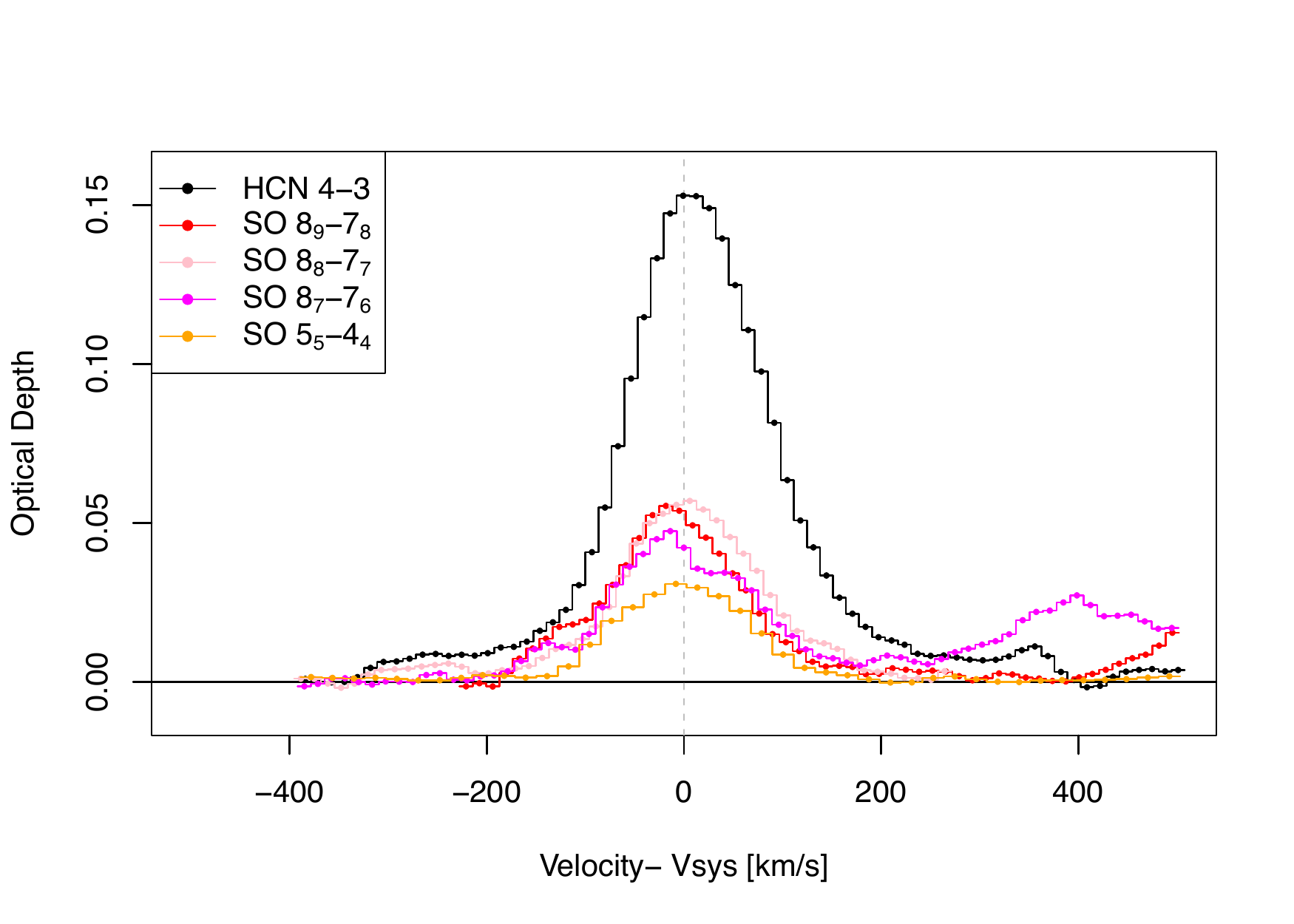}{0.475\textwidth}{(b)}}
\caption{(a) Optical depth profiles acquired in ALMA Band-3 and Band-4 observations. Black, purple, blue, and orange lines stand for HCN $J=1-0$, SO $J_N = 3_3 - 2_2$, SO $J_N = 3_2 - 2_1$, and SO $J_N = 2_2 - 1_1$ transitions, respectively. (b) Optical depth profiles of HCN and SO transitions in ALMA Band-6 and Band-7 observations presented in \citep{2020ApJ...895...73K}. The dashed vertical lines represent the systemic velocity.}
\label{fig:HCN_SO_opticalDepth}   
\end{figure}


Figure \ref{fig:HCN_SO_opticalDepth} shows optical depth profiles of HCN and SO absorption lines to be compared with higher frequency transitions acquired in ALMA Band-6 and Band-7 observations reported in \cite{2020ApJ...895...73K}.
While absorption profiles in Band-6 and Band-7 consist of a single component peaked at the systemic velocity with wider redward tails, those in Band-3 and Band-4 consist of multiple components with the redshifted peaks at $164 \pm 6$ km s$^{-1}$ and sharp redward edges around 210 km s$^{-1}$, and shallow blueward slopes down to $-190$ km s$^{-1}$.
The velocity ranges of Band-3 and Band-4 absorption features are settled within the HCN $J=4-3$ velocity range of $V_{\rm sys} \ ^{+382}_{-318}$ km s$^{-1}$.

We have measured equivalent width, EW$\displaystyle = \int \tau(v) dv$, by integrating the optical depth profile in the velocity range of $(-222, 258)$ km s$^{-1}$ after subtracting the line-free baseline in $(-312, -222) - (258, 478)$ km s$^{-1}$.
The standard error in EW is dominated by uncertainty in the baseline-level determination.
For SO $J_N = 3_3 - 2_2$, we corrected contamination of SO$_2$ $12_{2, 10} - 12_{1, 11}$ by assuming that the extra EW equals to the mean of those in $10_{2, 8} - 10_{1, 9}$ and $12_{2, 10} - 12_{1, 11}$.
We derived total column density, $N_{\rm tot}$, for SO, HCN, HCO$^+$, and CS assuming local thermodynamic equilibrium (LTE) as
\begin{equation}
N_{\rm tot} = \frac{3 k T_{\rm ex}}{8 \pi^3 \mu^2 B (2J+1)} \exp{\left(\frac{hB}{3kT_{\rm ex}}\right)} 
\exp{\left( \frac{hBJ(J+1)}{kT_{\rm ex}} \right)} 
\left[ \exp{\left( \frac{h\nu}{kT_{\rm ex}}\right)} -1 \right]^{-1} {\rm EW}, \label{eqn:columnDensity}
\end{equation}
where $k$ is the Boltzmann constant, 
$h$ is the Planck constant, 
$\mu$ is the permanent dipole moment of the molecule, 
$B$ is the rotational constant, and
$\nu$ is the rest frequencies.
We applied the excitation temperature, $T_{\rm ex} = 26$ K, obtained in section \ref{sec:temperature}.
The EW and $N_{\rm tot}$ values are listed in Table \ref{tab:absorptionLineID}.

\begin{deluxetable}{llrrrrr}
\tablecaption{Identified absorption lines}
\tablehead{
\colhead{Species} & \colhead{Transition} & \colhead{$\nu_{\rm rest}$} & \colhead{$\tau_{\rm max}$} & \colhead{Velocity} & \colhead{EW}            & \colhead{$N_{\rm tot}$} \\
\colhead{}        & \colhead{}           & \colhead{(GHz)}            & \colhead{}                 & \colhead{(km s$^{-1}$)}& \colhead{(km s$^{-1}$)} & \colhead{($10^{14}$ cm $^{-2}$)}
}
 \colnumbers
 \startdata
SO        & $J_N = 2_2 - 1_1$ &  86.09395 & $0.0079 \pm 0.0006$ & $165$ & $1.28 \pm 0.07$ & $2.33 \pm 0.13$ \\
          & $J_N = 3_2 - 2_1$ &  99.29987 & $0.0151 \pm 0.0009$ & $170$ & $2.33 \pm 0.11$ & $2.55 \pm 0.12$ \\
          & $J_N = 3_3 - 2_2$ & 129.13892 & $0.0214 \pm 0.0012$ & $155$ & $3.02 \pm 0.11$ & $2.47 \pm 0.09$ \\
          & $J_N = 4_5 - 4_4$ & 100.02964 & $0.0061 \pm 0.0012$ & $167$ & $0.72 \pm 0.11$ \\
          & $J_N = 8_8 - 7_8$ & 129.95366 & $0.0027 \pm 0.0008$ & $105$ & $0.23 \pm 0.07$ \\
          & $J_N = 13_{13} - 13_{12}$ & 139.50251 & $0.0030 \pm 0.0009$ & $229$ & $< 0.075$ \\
          & $J_N = 14_{14} - 14_{13}$ & 140.53709 & $0.0033 \pm 0.0011$ & $408$ & $0.15 \pm 0.05$ \\
          & $J_N = 15_{15} - 15_{14}$ & 141.41311 & $0.0013 \pm 0.0009$ & $482$ & $0.14 \pm 0.06$ \\
          & $J_N = 16_{16} - 16_{15}$ & 142.16005 & $0.0032 \pm 0.0011$ & $408$ & $< 0.09$ \\
SO$_2$    & $J_{Ka, Kc} = 6_{2,4} - 6_{1,5}  $   & 140.30617 & $0.0037 \pm 0.0011$ & $159$ & $< 0.07$ \\
          & $J_{Ka, Kc} = 10_{2,8} - 10_{1,9}$   & 129.51481 & $0.0060 \pm 0.0008$ & $161$ & $0.60 \pm 0.07$ \\
          & $J_{Ka, Kc} = 12_{1,11} - 11_{2,10}$ & 129.10583 & $-$ \\
          & $J_{Ka, Kc} = 12_{2,10} - 12_{1,11}$ & 128.60513 & $0.0059 \pm 0.0008$ & $164$ & $0.65 \pm 0.07$ \\
HCN       & $J = 1 - 0$ & 88.63185 & $0.0266 \pm 0.0007$ & $166$ & $4.95 \pm 0.07$ & $3.15 \pm 0.45$ \\
H$^{13}$CN & $J = 1 - 0$ & 86.33992 & $0.0045 \pm 0.0013$ &$-4$ & $0.64 \pm 0.10$& $0.43 \pm 0.07$ \\
HCO$^+$   & $J = 1 - 0$ & 89.18853 & $0.0128 \pm 0.0013$ & $149$ & $1.83 \pm 0.10$ & $0.68 \pm 0.05$\\
CS        & $J = 2 - 1$ & 97.98095 & $0.0050 \pm 0.0011$ & $17$ & $0.47 \pm 0.13$ & $0.41 \pm 0.11$ \\
\enddata
\tablecomments{ 
(1) Line species;  (2) Transition; (3) Rest frequency; (4) Peak optical depth; (5) Peak velocity with respect to the systemic velocity; (6) Equivalent width $= \int \tau dV$. The optical depth of SO$_2$ $J_{Ka, Kc} = 12_{1,11} - 11_{1,10}$ cannot be measured and it contaminates that of SO $J_N = 3_3 - 2_2$; (7) Total column density assuming $T_{\rm ex} = 26$ K.
}
\end{deluxetable} \label{tab:absorptionLineID}

\section{Discussion} \label{sec:Discussion}
Four species of sulfur-bearing molecules, H$_2$S, SO, SO$_2$, and CS, have been detected in absorption toward the nucleus of NGC 1052. We discuss about the origin of these molecules, physical conditions, and the heating mechanism.

\begin{figure}[ht!]
\plottwo{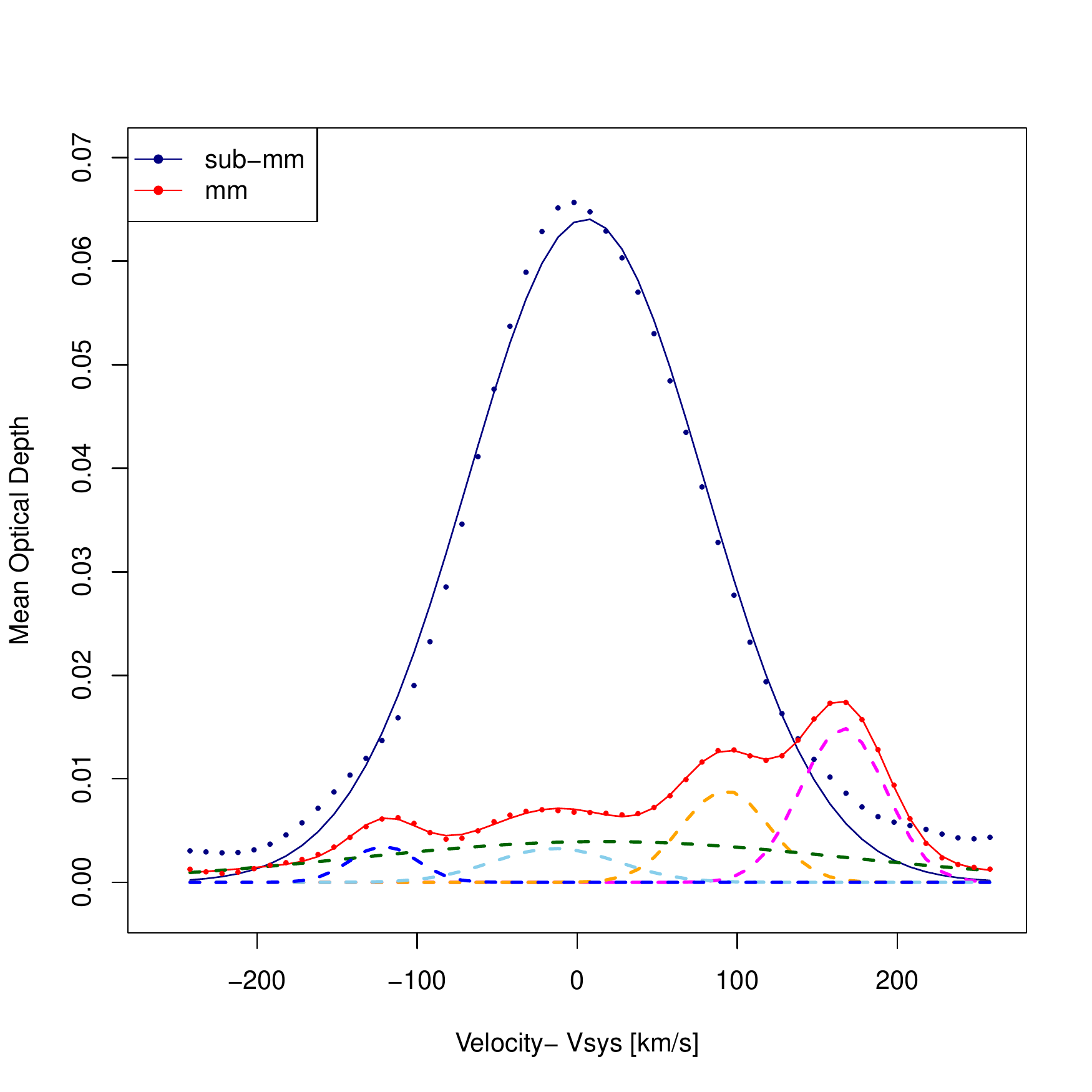}{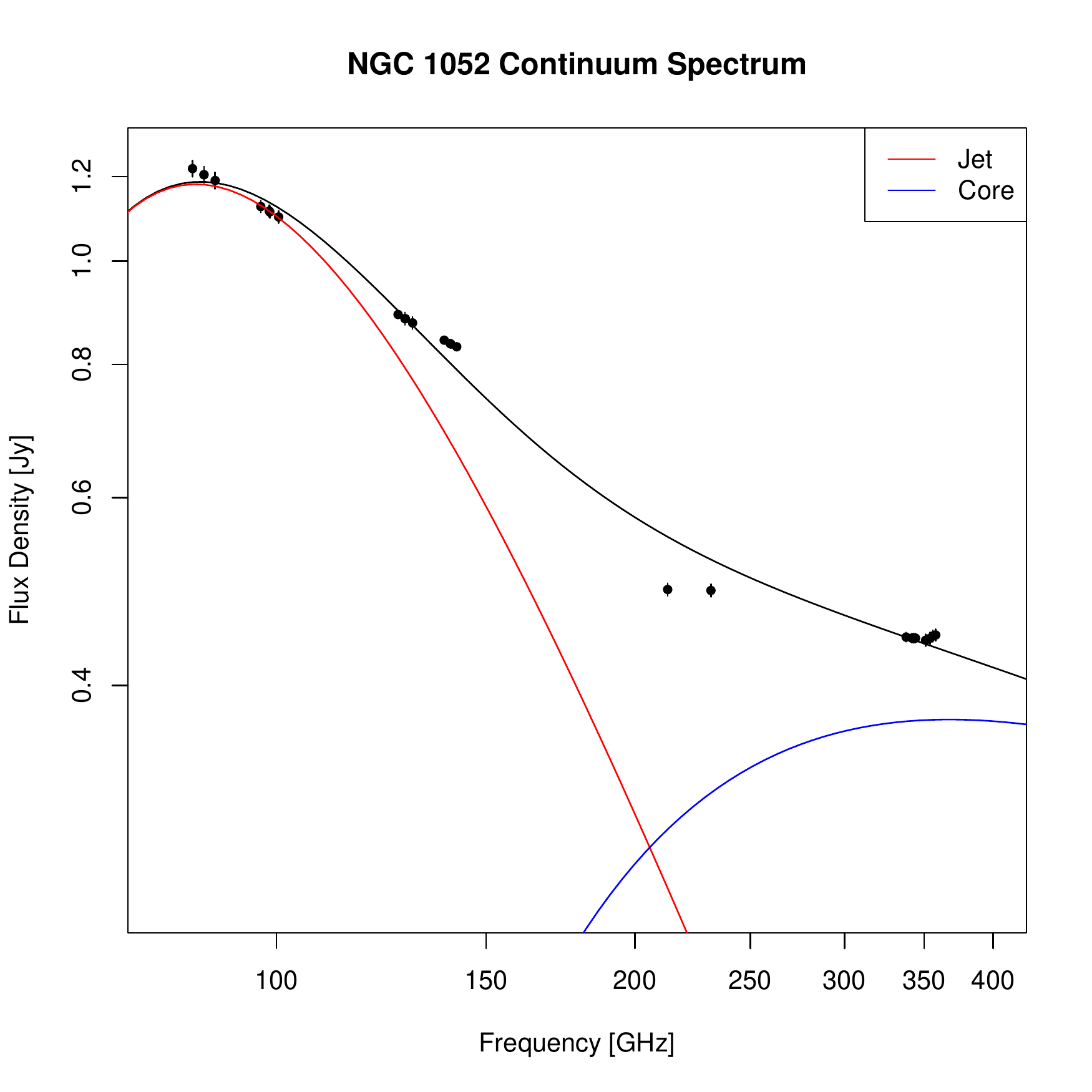}
\figcaption[OpticalDepthComponents.pdf]{Average optical depths in sub-mm and mm absorptions. Filled circles in navy blue and red stand for the average optical depths in sub-mm (HCN $J=4-3$, SO $J_N = 8_9 - 7_8, \ 8_8 - 7_7, \ 8_7 - 7_6,$ and $5_5 - 4_4$) and mm (HCN $J=1-0$, SO $J_N = 3_3 = 2_2, \ 3_2 - 2_1,$ and $2_2 - 1_1$) absorptions, respectively. Solid lines in navy blue and red indicate Gaussian fitting for the average optical depth in sub-mm (single component) and mm (5 components) absorptions, respectively. Gaussian components of the mm absorption are shown in dashed lines in magenta, orange, green, cyan, and blue. Parameters of the Gaussian components are listed in Table \ref{tab:GaussianComponent}\label{fig:OpticalDepthComponent}}
\figcaption[ContSpec.pdf]{A possible decomposition of the continuum spectrum of NGC 1052 in mm/sub-mm wavelengths. Black dots are ALMA measurements in this paper and \cite{2020ApJ...895...73K}. Jet (red) and core (blue) components are indicated by free--free absorbed synchrotron spectra with the spectral index of $-3$ and $-0.5$, respectively. Note that the decomposition is not unique but is tweaked to make the continuum in mm and sub-mm dominated by the jets and the core, respectively.\label{fig:ContSpec}}
\end{figure}


\begin{deluxetable}{lrrr}
\tablecaption{Gaussian Parameters of Absorber Components}
\tablehead{
\colhead{Component} & \colhead{Peak Opacity} & \colhead{Velocity} & \colhead{FWHM}  \\
\colhead{}     & \colhead{}          & \colhead{km s$^{-1}$}   & \colhead{km s$^{-1}$}   }
 \colnumbers
 \startdata
sub-mm (navy)  & $0.0642 \pm 0.0009$ &  $5.5 \pm 1.1$ & $174.3 \pm 2.7$  \\
mm (magenta)  & $0.0149 \pm 0.0004$ &  $166.1 \pm 0.6$ & $62.8 \pm 1.5$  \\
mm (orange)   & $0.0089 \pm 0.0008$ &  $92.5 \pm 0.9$ & $64.9 \pm 2.9$  \\
mm (green)    & $0.0039 \pm 0.0012$ &  $16.1 \pm 5.8$ & $360.0 \pm 49$  \\
mm (cyan)     & $0.0033 \pm 0.0011$ &  $-13.7 \pm 2.4$ & $92.1 \pm 17$  \\
mm (blue)     & $0.0035 \pm 0.0004$ &  $-121.0 \pm 1.0$ & $48.6 \pm 4.0$  \\
\enddata
\tablecomments{(1) Components with the color in Figure \ref{fig:OpticalDepthComponent}; (2) Peak optical depth; (3) Peak velocity with respect to the systemic velocity; (4) Velocity width in full width at half maximum.}
\end{deluxetable} \label{tab:GaussianComponent}

\subsection{Line profiles}
It is remarkable that optical depth profiles are classified into two distinct groups: (1) multi-component asymmetric profile peaked at redshifted velocity of $164 \pm 6$ km s$^{-1}$, presented in Figure \ref{fig:HCN_SO_opticalDepth}(a), and (2) single component peaked at the systemic velocity with wider redward tails, shown in Figure \ref{fig:HCN_SO_opticalDepth}(b).
Two groups are discriminated by the frequencies below 129 GHz and above 214 GHz.
Hereafter we name two groups ``mm absorption'' and ``sub-mm absorption''.

We averaged the optical depths for each group to highlight the difference and plotted in Figure \ref{fig:OpticalDepthComponent}.
The average of sub-mm and mm optical depths consist of the line species plotted in Figure \ref{fig:HCN_SO_opticalDepth}; sub-mm: HCN $J=4-3$, SO $J_N = 8_9 - 7_8, \ 8_8 - 7_7, \ 8_7 - 7_6,$ and $5_5 - 4_4$, and mm: HCN $J=1-0$, SO $J_N = 3_3 = 2_2, \ 3_2 - 2_1,$ and $2_2 - 1_1$.
While the sub-mm optical depth can be fitted by a single Gaussian component, the mm absorption requires at least 5 components. Parameters of the Gaussian components are listed in Table \ref{tab:GaussianComponent}. These components are also presented in Figure \ref{fig:OpticalDepthComponent}.
Velocities of the sub-mm and the mm green components coincide with the systemic velocity.
The mm absorption additionally contains two stronger redshifted and two weaker blueshifted components.
These redshifted and blueshifted components show significantly narrower velocity widths than that of the systemic velocity component.

\subsection{Temperature of the absorber} \label{sec:temperature}
\begin{figure}[ht!]
\plotone{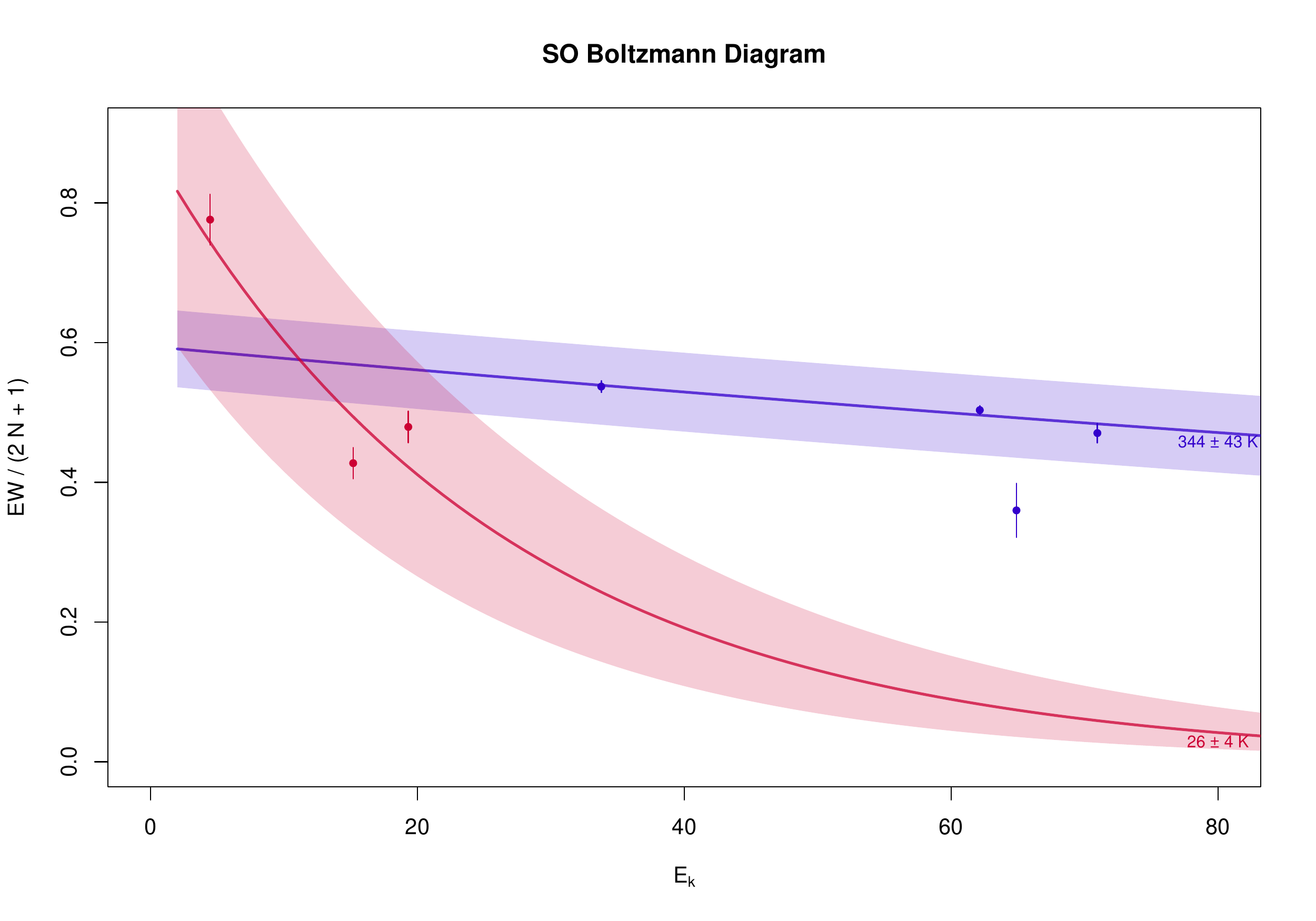}
\caption{Boltzmann diagram of SO absorption feature. Red and blue lines indicate the best fits for mm and sub-mm SO absorptions, respectively. The transparent bands show the uncertainties of the temperatures and intercepts.}
\label{fig:BoltzmannSO}
\end{figure}

We attempt to measure rotation temperature, $T_{\rm rot}$, by comparing EWs of SO in multiple transitions as
\[
\frac{\rm EW}{2N_l + 1} \sim n_0 \exp \left( - \frac{E_k}{T_{\rm rot}} \right),
\]
for these mm and sub-mm absorbers separately.
Here $N_l$ and $E_k$ are the lower-level rotational angular momentum and energy, respectively, and $n_0$ stands for the intercept.
The best-fit resulted in $T_{\rm rot} = 26 \pm 4$ K and $344 \pm 43$ K for the mm and sub-mm absorbers, respectively.
Figure \ref{fig:BoltzmannSO} shows the Boltzmann diagram of the detected permitted SO absorption lines.
SO $J_N = 3_3 - 2_2$, $3_2 - 2_1$, and $2_2 - 1_1$ transitions below 129 GHz are plotted in red markers. We added $5_5-4_4$, $8_7-7_6$, $8_8-7_7$, and $8_9-7_8$ transitions above 214 GHz in blue markers using data in \cite{2020ApJ...895...73K}.
To correct HC$^{15}$N contamination into $8_8-7_7$ feature, we applied proportional allocation of EW(SO):EW(HC$^{15}$N)$= 7.9:1.2$ using the result of two-component Gaussian decomposition.

The high temperature in sub-mm regime is consistent with the presence of vibrationally excited (v$_2$ = 1) HCN and HCO$^+$ lines \citep{2020ApJ...895...73K}.
Presence of SO$_2$ absorption features also supports the warm environment because SO$_2$ would froze onto dust particles under $T < 130$ K \citep{2020IAUS..345..360N}.
\cite{2020ApJ...895...73K} has discussed physical properties of the molecular torus, for the cases of $T=$ 50 K, 100 K, 230 K, and 500 K, to explain the obtained absorption line features.
Our results rule out the 50 K and 100 K cases, and is consistent with 230 K and 500 K that yields the covering factor of $f_{\rm cov} = 0.17^{+0.06}_{-0.03}$, the geometrical thickness of $\Theta = 0.4 - 1.0$, and the radius of $R = 2.4 \pm 1.3$ pc.
Non-detection of SiO features indicates that the temperature must be below the sublimation temperature of silicates.

The temperature of the mm absorber is below the freeze-out temperatures of H$_2$, SO and SO$_2$.
This indicates that sulfur-bearing molecules have been cooled after released from warm dust.
We will discuss about origin of cool mm absorber in section \ref{subsec:location} and \ref{subsec:molecularOrigin}.

\subsection{Location of absorbers} \label{subsec:location} 
Because an absorber must locate in front of the background continuum source along the line of sight, frequency-dependent continuum structure can cause the difference in absorption profiles.
While the continuum source is spatially unresolved in ALMA observations with the synthesized beam of $\sim 0^{\prime \prime}.3$, multifrequency VLBI studies have revealed sub-parsec-scale structures composed by the core and doble-sided jets.
The core is optically thick and hidden in the gap below 43 GHz due to free--free absorption \citep{1999AAS...194.2002K, 2001PASJ...53..169K, 2003PASA...20..134K, 2003A&A...401..113V, 2004A&A...426..481K}, and appears at higher frequencies with a peaked flat spectrum  \citep{2016A&A...593A..47B, 2016ApJ...830L...3S, 2019ApJ...872L..21S}.
On the other hand, the jets have a steep spectrum with a brightness gradient inward the core \citep{2001PASJ...53..169K, 2003PASA...20..134K, 2003A&A...401..113V, 2004A&A...426..481K, 2016A&A...593A..47B, 2019A&A...623A..27B} with a peak in GHz frequencies \citep{2019A&A...629A...4F}.
As a result, the extension of the jet becomes longer at lower frequency.
The footpoints of the jets are optically thick below 22 GHz with a gap due to the free--free absorption.
Although no VLBI image at sub-mm has been publicized to date, the sub-mm continuum emission is putatively dominated by the core component rather than the jet.
Figure \ref{fig:ContSpec} shows a possible decomposition of the continuum spectrum in mm/sub-mm wavelengths measured by ALMA.
This decomposition is not unique but is tweaked to make the mm and sub-mm continuum are dominated by the jet and the core, respectively.

KVN observations at 86 GHz has revealed that HCN and HCO$^+$ absorbing clumps are smaller than 0.1 pc and locate on the western receding jet 0.24 pc -- 0.27 pc offset from the continuum peak implying complex kinematics in the vicinity of the active galactic nucleus, such as inflow, outflow, turbulence \citep{2016ApJ...830L...3S, 2019ApJ...872L..21S}.
The HCN absorption is dominated by two redshifted components at 149 km s$^{-1}$ and 212 km s$^{-1}$ which represent the redshifted component in the ALMA results. The blueshifted component with the peak opacity of 0.0035 measured ALMA was not identified with KVN where rms noise level was 0.035 in the normalized spectrum and bandwidth of 128 MHz was not sufficient to determine line-free baseline \citep{2016ApJ...830L...3S}.
The KVN HCO$^+$ absorption profile with FWHM of $272\pm 50$ km s$^{-1}$ covers the systemic-velocity \citep{2019ApJ...872L..21S} that is more consistent with the ALMA mm absorption.
While the western receding jet casts both redshifted and blueshifted HCO$^+$ absorptions, the eastern approaching jet does not show clear absorption feature. Higher dynamic-range VLBI observation is necessary to reproduce the ALMA results. 

\begin{figure}
\plotone{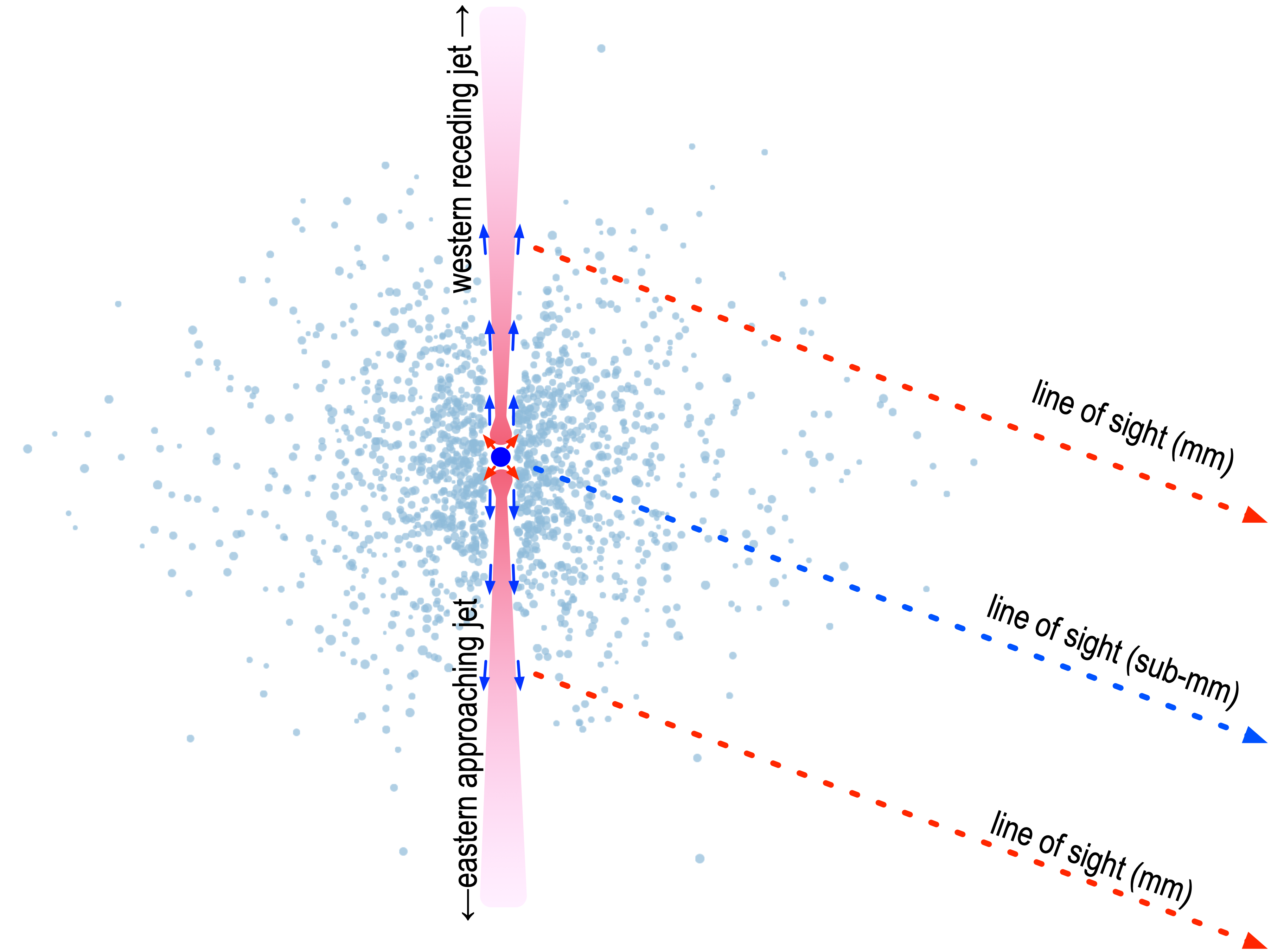}
\figcaption[figures/ClumpyTorus.png]{Schematic diagram of the torus composed by molecular/dusty clumps (pale blue). The core (blue circle) contains the central engine and the jet-launching region with sub-mm flat-spectrum synchrotron emission. The core is hidden in low frequency regime due to free--free absorption in the inner area of the torus. High pressure outflows (red arrows) from the core interact with the inner wall of the torus. HCN, HCO$^+$, and sulfur-bearing molecular gas evaporate from dust grains in the torus by the shock and produce sub-mm absorption features. The reverse shock collimates the jet along the polar axis. The jet emits optically thin synchrotron radiation off the torus. The jet drags molecular gas downstreams (blue arrows) which produce mm absorption features. The line of sight toward the western receding side has greater optical depth than the eastern approaching side to generate the asymmetric line profiles.\label{fig:ClumpyTorus}}
\end{figure}

In view of the characteristics of the continuum components, mm and sub-mm absorbers are expected to locate where they mainly cover the jet and the core, respectively, as illustrated in Figure \ref{fig:ClumpyTorus}.
This model predicts longer line of sight toward the western receding side than that toward the eastern approaching side to generate the asymmetric line profiles, with a greater optical depth in redshifted component, as is observed in mm abosorption.
The model interprets the velocity width of the sub-mm absorption as the turbulent motion.
In addition, the redshifted and blueshifted mm abosorptions represent dragged gas flow by the approaching and recedig jets, respectively.

Alternative explanation for the redshifted and blueshifted mm abosorptions is that clumps inside the torus are orbiting the core and pass across the line of sight to the continuum emission.
In this case, the mm absorber is expected to locate at further distance from the core than the sub-mm absorbers to explain the lower excitation temperature.
Since the jets are well collimated and are expected to align with the orbital axis, non-circular orbit is required to produce the redshifted and blueshifted absorption profiles.

\subsection{Origin of sulfur-bearing molecules} \label{subsec:molecularOrigin}
In protoplanetary disks, sulfur-bearing molecules are known to trace evaporation of icy components of dust. \cite{2020IAUS..345..360N} calculated evolution of fractional abundances of H$_2$S, SO, SO$_2$, and CS under given temperature and demonstrated that evaporated H$_2$S is destroyed by gas-phase reactions to form SO and SO$_2$, and these  will be frozen out on grains if dust temperature is colder than 60 K and 130 K, respectively.

The SO-to-CS abundance ratio in protostars spreads in various ranges; $N({\rm SO})/N({\rm CS}) < 0.11$ in the solar-analogue IRAS 16293-2422 B \citep{2018MNRAS.476.4949D}, $1.7 – 37.5$ in 15 positions of molecular shocks in the low mass protostar NGC 1333 IRAS 2 \citep{2005A&A...437..149W}, $1.1 - 1.9$ in the shocked region of class 0 protostar L 1157 outflow\citep{1997ApJ...487L..93B}, and $3^{+13}_{-2}\times 10^2$ in the  Class I protostellar source Elias 29 \citep{2019ApJ...881..112O}.
The ratio in NGC 1052 is $6.1 \pm 0.28$ and $2.55 \pm 0.05$ for mm and sub-mm\footnote{Sub-mm column densities are derived using EW values of SO $5_5 - 4_4$, $8_7 - 7_6$, $8_8 - 7_7$, and $8_9 - 7_8$ transitions and CS $7 - 6$, HCN $4-3$, HCO$^+$ $4-3$ in \citet{2020ApJ...895...73K} and applying $T_{\rm ex} = 344$ K.} absorptions, respectively, which are within the range of protostars and matches to molecular shocks in NGC 1333 IRAS 2.

\begin{deluxetable}{lrrrrrl}
\tablecaption{Integrated line intensity}
\tablehead{
\colhead{Galaxy} & \colhead{HCO$^+$ $(4 - 3)$} & \colhead{SO $(8_8 - 7_7)$} & \colhead{HCN $(1 - 0)$} & \colhead{HCO$^+$ $(1 - 0)$} & \colhead{SO $(3_2 - 2_1)$} & Ref.  \\
\colhead{}       & \colhead{Jy km s$^{-1}$}  & \colhead{Jy km s$^{-1}$} & \colhead{Jy km s$^{-1}$} & \colhead{Jy km s$^{-1}$}     & \colhead{Jy km s$^{-1}$}    &  }
 \colnumbers
 \startdata
IRAS 20551-4250  & $16.9 \pm 0.2$            &  $0.64\pm 0.15$  \\
NGC 1068         &                           &                          &  $26.4\pm 0.6$        & $13.3\pm 0.7$             & $0.8\pm 0.2$ & 1 \\
NGC 253          &                           &                          &  $70\pm 1$            & $59\pm 1$                 & $4.2\pm 0.3$ & 2 \\
IC 342           &                           &                          &  $11.6\pm 0.1$        & $8.4\pm 0.1$              & $0.9\pm 0.1$ & 2 \\
NGC 1052 (EW: km s$^{-1}$) &  $15.9 \pm 0.4$ &  $7.9\pm 0.5$            &  $4.95\pm 0.07$       & $1.83\pm 0.13$            & $2.33\pm 0.11$ & 3, 4\\
\enddata
\tablecomments{References: 1.\citet{2017ApJ...849...29I}, 2.\citet{2019PASJ...71S..20T}, 3.\citet{2020ApJ...895...73K}, 4.this work}
\end{deluxetable} \label{tab:IntegratedIntensity}

The ratios of SO-to-HCN and SO-to-HCO$^+$ integrated line intensities in nearby galaxies are summarized in Table \ref{tab:IntegratedIntensity}.
NGC 1052 shows significantly higher SO-to-HCN and SO-to-HCO$^+$ ratios than does other galaxies.
SO ($8_8 - 7_7$):HCO$^+ \ (4-3) = 1:26$ and $1:3$ in the ultraluminous infrared galaxy, IRAS 20551-4250 and NGC 1052.
SO $(3_2-2_1)$:HCN $(1-0)$:HCO$^+ \ (1-0) = 1:33:17$, $1:17:14$, $1:13:9$, and $1:2:0.8$ in the Seyfert galaxy NGC 1068, in the starbust galaxies NGC 253 and IC 342, and NGC 1052, respectively.

The SO-to-HCO$^+$ abundance ratio in NGC 1052 sub-mm absorption is derived to $4.63 \pm 0.03$ applying $T_{\rm ex} = 344$ K.
This value is probably overestimated because HCO$^+$ is optically thick as implied by high H$^{13}$CN-to-H$^{12}$CN ratio \citep{2020ApJ...895...73K}. The ratio will be $0.95 \pm 0.03$ after applying correction with the $^{12}$C-to-$^{13}$C abundance ratio of $\sim 50$ and the covering factor of $f_{\rm cov} = 0.17$.
That is significantly higher than the SO-to-HCO$^+$ abundance ratio of $0.35 \pm 0.23$ in IRAS 20551-4250, applying $T_{\rm ex} =36$ K \citep{2017ApJ...849...29I}.

The SO-to-HCN abundance ratio in mm abosorption is derived to $0.78 \pm 0.03$, with $T_{\rm ex} = 26$ K.
This value will be corrected to $0.07 \pm 0.02$ by the $^{12}$C-to-$^{13}$C abundance ratio and the covering factor.
Although the excitation temperatures in NGC 1068, NGC 253, and IC 342 are unknown, $T_{\rm ex} = 20 - 500$ K yields the SO-to-HCN abundance ratio of $0.04 - 0.06$, $0.08 - 0.11$, and $0.11 - 0.14$, respectively \citep{2019PASJ...71S..20T}.
Thus, the opacity-corrected SO-to-HCN abundance ratio of mm absorption in NGC 1052 is comparable to that in other nearby galaxies.

The torus in NGC 1052 is considered to be dust rich as a reservoir of sulfur-bearing molecules.
\cite{1999ApJ...515L..61B} revealed polarized broad H$\alpha$ emission indicating presence of dust torus which obscures direct light from the broad line region (BLR). The BLR light would escape only in polar-axis direction and scattered light can be observed. Thus, it is natural to consider presence of dust grains in the molecular torus.
High SO-to-HCO$^+$ abundance ratio in sub-mm absorption is interpreted by evaporation of sulfur-bearing molecules from dust grains under high temperature $>130$ K.

The mm absorber is cooler than the freeze-out temperature, contains redshifted and blueshifted velocity components up to $166$ km s$^{-1}$ with a narrow width of $<65$ km s$^{-1}$, and the SO-to-HCN abundance ratio is comparable to other nearby galaxies.
A plausible interpretation for the redshifted and blueshifted mm absorber is that they are downstreams of sub-mm absorber dragged by the jets.
Another possibility is that orbiting clumps pass across the line of sight toward mm continuum source by chance.
High-dynamic-range VLBI monitoring is desired to clarify the location and dynamics of the mm and sub-mm absorbers.


\subsection{The heat source}
Dust grains with a radius of $r$ emit blackbody radiation of
\begin{eqnarray}
P_{\rm rad, out} = 4\pi r^2 \sigma_{\rm s} T^4, \label{eqn:coolingPower}
\end{eqnarray}
where $\sigma_{\rm s}$ is the Stefan--Boltzmann constant.
If we apply $T = 344$ K and $r = R = 2.4$ pc (see section \ref{sec:temperature}), the radiation power of the whole torus will be $P_{\rm rad, out} = 5.5 \times 10^{37}$ W which exceeds the bolometric luminosity by two orders of magnitude.
Therefore, the high temperature must appear only in small fraction of the torus body.
The radiation quickly cool the gas in the torus unless presence of a heat source.
In this section we attempt to examine possible heating mechanism to maintain such a high temperature of gas and dust in the torus.

\subsubsection{Jet--torus interaction}
NGC 1052 emanates sub-relativistic double-sided jets with the bulk speed of $0.26c - 0.53c$ \citep{2003A&A...401..113V}.
The width of the jets is strongly confined to 0.016 pc ($1.3 \times 10^3 \ R_{\rm s}$) inside $10^4 \ R_{\rm s}$ from the nucleus \citep{2020AJ....159...14N,2022A&A...658A.119B}, implying pressure gradient of the torus.
Interaction between the relativistic jets and the confining molecular torus can cause shock that heats gas and dust.

Kinetic power of the jets is estimated as 
\begin{eqnarray}
P_{\rm jet, tot} = \frac{1}{2} \pi r^2_j \rho_j v^3_j, \label{eqn:jetpower}
\end{eqnarray}
where $r_j = 0.008$ pc is the radius of cylindrical jets, $\rho_j$ is the density, and $v_j = 0.26c$ is the jet speed.
At the shock front between the torus gas and jets with the over-pressure factor of $\sim 1.5$ \citep{2019A&A...629A...4F}, we have
\begin{eqnarray}
\rho_j  v^2_j = 1.5 \rho_t v^2_s, \label{eqn:shockBalance}
\end{eqnarray}
where $v_s$ is the advancing speed of the shock. The mean torus gas density, $\rho_t$, has been estimated as $n_{\rm H2} = 4.4 \times 10^6$ cm$^{-3}$ \citep{2020ApJ...895...73K}.
Giving $v_s \simeq \Delta V_{\rm wing} = 350$ km s$^{-1}$ and $v_j = 0.26c$, we have $\rho_j = 4.4 \times 10^{-19}$ kg m$^{-3}$ and $P_{\rm jet, tot} = 2.0 \times 10^{34} \ {\rm W}$, which corresponds to 10\% of the bolometric luminosity of AGN.

For the radiative cooling (equation \ref{eqn:coolingPower}) localized in the shock-heating region of $r_j$, we have $P_{\rm rad, out} = 6 \times 10^{32}$ W or 3\% of the kinetic power of the jets.
Therefore, the jet--torus interaction can account for the heating of gas and dust with a fractional amount of jet power.
Because this model presumes localized heating within collimated jet width, it requires less energy injection than heating whole the torus.

Another advantage of this model is to naturally explain the evaporation of sulfur-bearing molecules from dust mantle via shocks generated by jet--torus interaction, and presence of cool downstream gas to show mm absorption features.
Future mm/sub-mm VLBI spectral images would unveil temperature distribution of sulfur-bearing molecules along the jets to clarify the cooling mechanism.

In following subsections we consider other mechanism for heating the gas in which the SO feature reside.

\subsubsection{AGN radiation}
Here we examine if the bolometric luminosity of $L_{\rm bol} = 2 \times 10^{35}$ W would be enough to maintain the warm environment in the torus.
A dust grain with a radius of $r$ at the distance $R=2.4$ pc from the nucleus receives radiation power of $\displaystyle P_{\rm rad, in} = L_{\rm bol} \frac{r^2}{4R^2}$.
Under the balance $P_{\rm rad, in} = P_{\rm rad, out}$, we have $L_{\rm bol} = 16 \pi R^2 \sigma_{\rm s} T^4$ and then $T= 60$ K.
This indicates that AGN radiation cannot account for heating the gas and dust in the 2.4-pc torus.
Radiative heating to 344 K would be realized at closer to the nucleus by factor of 33, i.e., $R \sim 0.07$ pc.

\subsubsection{Accretion}
Accretion matter converts its potential energy into kinetic energy by 50\% and dissipates by 50\%.
Here we examine if dissipation would account for the heating of gas and dust heating.

Under the gravity of $\displaystyle F_{\rm G} = \frac{GM_{\rm BH}}{R^2}$, dissipation rate for a unit mass will be $\displaystyle \frac{1}{2} F_{\rm G} v_{\rm acc}$, where $v_{\rm acc}$ is the accretion velocity that must be less than the absorption-line velocity width $\Delta V_{\rm wing} = 350$ km s$^{-1}$. 
For a dust grain with the mass of $\displaystyle m_{\rm dust} = \frac{4\pi}{3}\rho r^3$, the dissipation power is estimated as 
\begin{eqnarray}
P_{\rm acc, in} = \frac{2\pi}{3} \frac{GM_{\rm BH}}{R^2} \rho r^3 v_{\rm acc}. \label{eqn:accretionPower}
\end{eqnarray}
Comparing equation \ref{eqn:accretionPower} with the radiative cooling (equation \ref{eqn:coolingPower}) for the dust grain, we have
\[
\frac{P_{\rm acc, in}}{P_{\rm rad, dust}} = \frac{GM_{\rm BH} \rho r v_{\rm acc}}{6\sigma_s R^2 T^4} = 3 \times 10^{-7} \left( \frac{M_{\rm BH}}{1.5 \times 10^8 \ {\rm M}_{\odot}} \right) \left( \frac{\rho}{10^3 \ {\rm kg \ m}^{-3}} \right) \left( \frac{R}{2.4 \ {\rm pc}} \right)^{-2} \left( \frac{T}{344 \ {\rm K}} \right)^{-4} \left( \frac{r}{1 \ \mu {\rm m}} \right).
\]
Thus, accretion power would be a significant heating process at much inner region of $\sim 0.001$ pc but cannot account for the gas and dust heating at $R=2.4$ pc.

\subsubsection{Star formation}
Starburst activity in a circumnuclear disk or a molecular torus can be the origin of heat source in some Seyfert galaxies .
Starbust characteristics in the disk or clumpy torus can be diagnosed by the Toomre's $Q$ parameter defined as $\displaystyle Q = \frac{\sigma_R \kappa}{\pi G \Sigma}$, where $\sigma_R$ is the velocity width, $\kappa$ is the epicyclic frequency, and $\Sigma$ is the surface density \citep{2008A&A...491..441V, 2016ApJ...827...81I}.

As shown in \cite{2020ApJ...895...73K}, the circumnuclear disk in NGC 1052 is too gas-poor to drive mass accretion via ongoing star formation.
For the molecular torus, we estimate $Q \sim 8$, applying $\displaystyle \Sigma = \frac{M_{H_2}}{\pi R^2} = 7.2 \times 10^5$ $M_{\odot}$ pc$^{-2}$, the velocity width of $\sigma_R = 350$ km s$^{-1}$, and $\kappa = V_{\rm rot}/R = 8.1 \times 10^{12}$ rad s$^{-1}$.
Thus, it is classified as a massive, opaque, collisional disk according to \cite{2008A&A...491..441V} where a star formation rate does not play a significant role compared with cloud collisions.

\section{Conclusions} \label{sec:conclusion}
We have probed absorption features of sulfur-bearing molecules, H$_2$S, SO, SO$_2$, and CS, toward mm and sub-mm continuum emission in the radio galaxy NGC 1052, and obtained clues on physical properties of the molecular torus as summarized below.

\begin{enumerate}
\item Equivalent widths of SO absorption is comparable to HCN and HCO$^{+}$. After correcting optical depth and covering factor, sub-mm abosorption shows significantly high SO-to-HCO$^+$ abundance ratio than does the ultra-luminous infrared galaxy, IRAS 20551-4250. The SO-to-HCN abundance ratio in mm absorption is comparable to that in nearby starburst and Seyfert galaxies. The SO-to-CS abundance ratio matches to molecular shocks in the low mass protostar NGC 1333 IRAS 2.

\item While sub-mm HCN and SO absorption profiles are simple single component centered at the systemic velocity, mm absorption is multi-component asymmetric profile with a peak and sharp edge in redward and a shallower blueward slope. The difference of the profiles is ascribed to the location of the background continuum, the core and the jets in sub-mm and mm regimes, respectively.

\item The sub-mm SO absorption indicate the rotation temperature of $T_{\rm rot} = 344 \pm 43$ K. That is greater than the freeze-out temperatures of SO$_2$ and SO, and is consistent with the presence of 22-GHz H$_2$O masers and vibrationally excited HCN and HCO$^+$ absorption lines in the molecular torus.

\item Jet-torus interaction is the most plausible heat source for the gas and dust in the torus. Our results support that the torus is working for jet collimation inside 10$^4 \ R_{\rm s}$.

\item The mm SO absorption features fit $26 \pm 4$ K, significantly cooler than those in sub-mm. The mm continuum is dominated by jets and the cooler SO gas are considered to be dragged downstream flows. The asymmetric mm absorption profiles can be generated by different line of sights through the absorbers toward the western receding side and eastern approaching side.

\end{enumerate}

Following the mm/sub-mm absorption study with ALMA drew jet-torus interaction, we desire to acquire spatially resolved distribution of temperature and velocity field of the absorber using mm/sub-mm spectral VLBI observations.

\bigskip
We thank the anonymous reviewer for thorough and constructive suggestions.
This paper makes use of the following ALMA data: ADS/JAO.ALMA\#2013.1.01225.S and ADS/JAO.ALMA\#2016.1.00375.S. ALMA is a partnership of ESO (representing its member states), NSF (USA) and NINS (Japan), together with NRC (Canada), MOST and ASIAA (Taiwan), and KASI (Republic of Korea), in cooperation with the Republic of Chile. The Joint ALMA Observatory is operated by ESO, AUI/NRAO and NAOJ.
This work is supported by JSPS KAKENHI 18K03712 and 21H01137. NK is supported by JSPS KAKENHI 19K03918. 
KK acknowledges the support by JSPS KAKENHI Grant Number 17H06130.
DE acknowledges support from: (1) a Beatriz Galindo senior fellowship (BG20/00224) from the Spanish Ministry of Science and Innovation, (2) projects PID2020-114414GB-100 and PID2020-113689GB-I00 financed by MCIN/AEI/10.13039/501100011033, (3) project P20\_00334  financed by the Junta de Andaluc\'{i}a, and (4) project A-FQM-510-UGR20 of the FEDER/Junta de Andaluc\'{i}a-Consejer\'{i}a de Transformaci\'{o}n Econ\'{o}mica, Industria, Conocimiento y Universidades.

\vspace{5mm}
\facilities{ALMA}
\software{CASA 6.4.0.16}

\bibliography{NGC1052SOrev1}{}
\bibliographystyle{aasjournal}
\end{document}